\documentclass[twocolumn,prb,superscriptaddress]{revtex4-2}
\usepackage{amsfonts,amssymb,amsmath}
\usepackage{graphicx}
\usepackage{bbm}
\usepackage{mathcomp}
\usepackage{comment}
\usepackage[colorlinks,citecolor=blue]{hyperref}

\begin{document}
	\title{Anisotropic dynamical reconstruction of quantum geometry in quenched Chern insulators}

\author{Nan Li}
	\affiliation{State Key Laboratory of Quantum Optics Technologies and Devices, Institute of Laser Spectroscopy, Shanxi University, Taiyuan 030006,
		China} 
	\affiliation{Collaborative Innovation Center of Extreme Optics,
		Shanxi University, Taiyuan 030006, China}	
	  				
\author{Zhiyong Liu}
	\affiliation{State Key Laboratory of Quantum Optics Technologies and Devices, Institute of Laser Spectroscopy, Shanxi University, Taiyuan 030006,
		China} 
	\affiliation{Collaborative Innovation Center of Extreme Optics,
		Shanxi University, Taiyuan 030006, China}	
		
\author{Zhaoxin Liang}\thanks{zhxliang@zjnu.edu.cn}
	\affiliation{Department of Physics, Zhejiang Normal University, Jinhua, 321004, China}

\author{Ying Hu}\thanks{huying@sxu.edu.cn}
	\affiliation{State Key Laboratory of Quantum Optics Technologies and Devices, Institute of Laser Spectroscopy, Shanxi University, Taiyuan 030006, China} 
	\affiliation{Collaborative Innovation Center of Extreme Optics,
		Shanxi University, Taiyuan 030006, China}

\begin{abstract}
Under unitary dynamics, the Chern number of an evolving quantum state remains conserved even when a quench drives the Hamiltonian across a topological phase transition. In sharp contrast, we reveal an anisotropic dynamical reconstruction of the quantum metric. Following a sudden quench in a two-dimensional Chern insulator, the metric develops a principal frame in which one eigenvalue grows in time whereas the other remains nearly unchanged. At long times, the associated axes align with the energy-gradient direction and the tangent direction of the constant-energy contours of the post-quench Hamiltonian, respectively. This dynamically selected frame is distinct from that of the static post-quench ground-state metric. We identify momentum-dependent relative dynamical phases as its origin: they enhance the distinguishability of neighboring states separated along the energy-gradient direction, while states along an equal-energy contour remain nearly phase locked. Consequently, local and global metric observables acquire characteristic long-time signatures of the post-quench Hamiltonian. Our results establish a nonequilibrium mechanism by which coherent dynamics reorganizes quantum geometry, suggesting new possibilities for its dynamical control.
\end{abstract}

\maketitle

\section{Introduction}

The geometry of quantum states provides a fundamental framework for understanding quantum matter beyond conventional band dispersion~\cite{Torma2023,Liu2025,Yu2025}. The quantum geometric tensor (QGT)~\cite{Provost1980, Ma2010} unifies the local geometry of Bloch wave functions in parameter space. Its antisymmetric part, the Berry curvature, governs geometric phases and underlies topological invariants such as the Chern number~\cite{Berry1984,Xiao2010}. Its symmetric part, the quantum metric~\cite{Fubini1904,Study1905}, quantifies the distinguishability of nearby quantum states. It has emerged as an essential ingredient in a broad range of phenomena and phases of matter, including flat-band
superfluidity~\cite{Peotta2015,Hu2019,Xie2020,Julku2021}, nonlinear optical responses~\cite{Ahn2022,Li2026}, fractional Chern
insulators~\cite{Roy2014,WuL2024,Shavit2024}, and excitonic condensates~\cite{Hu2022}. Experimental detections of
quantum metric in ultracold atoms~\cite{Asteria2019}, superconducting circuits~\cite{Tan2019}, NV centers~\cite{Yu2020}, and solid-state systems~\cite{Kim2025,Zhao2025} have recently been reported. 

Despite this progress, quantum geometry has been largely restricted to Bloch wavefunctions of static Hamiltonians. A crucial yet much less explored question concerns the geometry of a coherently evolving quantum state, whose wavefunction continuously evolves under unitary dynamics and can stay out of equilibrium. Unlike a static eigenstate, a coherently evolving quantum state encodes the interplay between the initial condition, the driving Hamiltonian, and the momentum-dependent dynamical phases accumulated during evolution. While recent works have begun to explore nonequilibrium quantum geometry~\cite{Davide2020,Wu2024,Tang2026}, understanding how these dynamical ingredients reorganize the quantum geometry remains an open question.

Quantum quenches in Chern insulators provide a natural setting to address this question. Previous theoretical and experimental studies~\cite{Alessio2015,Hu2016,Caio2016,Wilson2016,Sun2018} have shown that, even when a quench drives the Hamiltonian across a topological phase transition, the Chern number of the evolving state remains conserved under unitary dynamics. This sharp separation between the topology of the state and that of the post-quench Hamiltonian motivates us to study whether the quantum metric retains initial-state memory or is dynamically reorganized by coherent evolution.

In this work, we investigate the quantum geometry of coherently evolving states following sudden quenches in the paradigmatic two-dimensional (2D) massive Dirac model. We analyze the quantum metric at both local and global levels. In contrast to the conserved Chern number, we find the 2D metric develops a pronounced anisotropy under unitary evolution: one principal component grows in time, whereas the orthogonal component remains nearly unchanged. We show that this behavior originates from the momentum dependence of the relative dynamical phase accumulated after the quench. States with different post-quench energies acquire an increasing relative phase and become more distinguishable, while states on the same constant-energy contour remain nearly phase locked. This mechanism selects a principal-axis structure tied to the energy-gradient and equal-energy directions of the post-quench Hamiltonian, distinct from the metric structure of its static ground state. It also gives rise to characteristic long-time momentum-space distributions of the quantum metric and to distinct dynamical behaviors of the quantum volume and gauge-invariant Wannier spread. Thus, in the 2D quenched Chern insulators considered here, coherent dynamics produces an anisotropic reconstruction of the quantum metric: the dynamically amplified component is controlled by the post-quench energy dispersion, whereas the subleading component retains memory of the initial state.

The remainder of this paper is organized as follows. In Sec.~\ref{sec2}, we briefly review the quantum geometry of the static massive Dirac model. In Sec.~\ref{sec3}, we investigate the dynamical evolution of both local and global quantum geometric quantities following a sudden quench. In Sec.~\ref{sec4}, we demonstrate the anisotropic dynamical reorganization of the quantum geometry.  Finally, Sec.~\ref{sec5} summarizes our main conclusions and discusses potential experimental realizations.

\begin{figure}[tb]
\centering
\includegraphics[width=1\columnwidth]{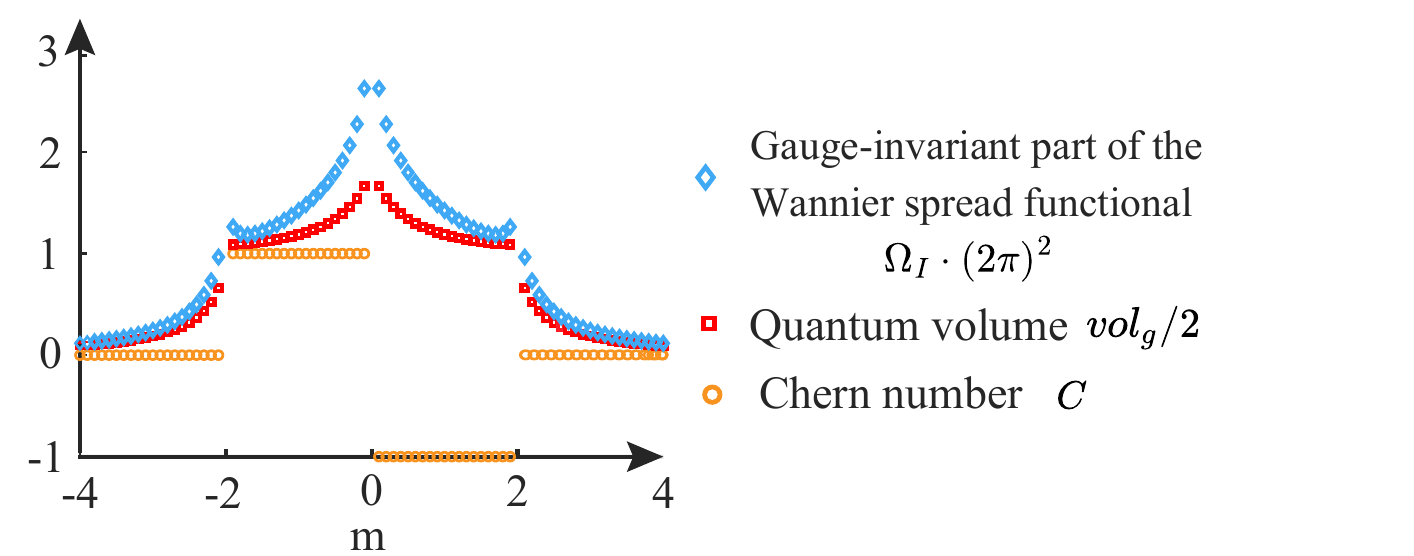}
\caption{Quantum geometry in a static Chern insulator. The Chern number $C$, quantum volume $\mathrm{vol}_g$, and gauge-invariant Wannier spread functional $\Omega_I$ of the ground state of the massive Dirac model are shown for various mass $m$.}
\label{Fig1}
\end{figure}

\section{Background: Quantum geometry of a Static Chern insulator}\label{sec2}

Before turning to the quench dynamics, we briefly review the quantum geometry of a static two-dimensional Chern insulator. At lattice momentum \(\mathbf{k}\equiv(k_x,k_y)\), the Bloch Hamiltonian takes the generic form $
H(\mathbf{k})=\mathbf{d}(\mathbf{k})\cdot\boldsymbol{\sigma}$,
where $\boldsymbol{\sigma}$ denotes the Pauli matrices, and $\mathbf d(\mathbf k)=d(\mathbf k)\hat{\mathbf d}(\mathbf k)$. Here $\hat{\mathbf d}(\mathbf k)$ specifies the Bloch-sphere texture, while $\pm d(\mathbf{k})=\pm E(k)$ corresponds to the band dispersion. Denoting the ground state of $H(\mathbf{k})$ by $|\psi({\bf k})\rangle$, its QGT is~\cite{Provost1980,Yu2025,Liu2025}:
\begin{eqnarray}
Q_{\mu\nu}(\mathbf{k}) &=& \langle \partial_{k_\mu} \psi({\bf k})| \left(\mathbb{I} - |\psi(\mathbf{k})\rangle\langle\psi(\boldsymbol{\bf k}) |\right)|\partial_{k_\nu} \psi(\boldsymbol{\bf k}) \rangle\nonumber\\
&=&G_{\mu\nu}(\mathbf{k}) - \frac{i}{2} B_{\mu\nu}(\mathbf{k}), \label{eq0}
\end{eqnarray}
 where $\mu, \nu \in \{x, y\}$ and $\mathbb{I}$ denotes the identity matrix. In Eq.~(\ref{eq0}), the real symmetric part is the quantum metric $G_{\mu\nu}(\mathbf{k})$, which measures the distinguishability between neighboring quantum states, while the imaginary antisymmetric part is the Berry curvature $B_{\mu\nu}(\mathbf{k})$. For a two-band model,
\begin{eqnarray}
G_{\mu\nu}(\mathbf{k}) & =&\frac{1}{4} \, \partial_{k_\mu} \hat{\mathbf{d}}(\mathbf{k}) \cdot \partial_{k_\nu} \hat{\mathbf{d}}(\mathbf{k}), \nonumber\\
B_{\mu\nu}(\mathbf{k}) & =& \frac{1}{2} \, \hat{\mathbf{d}}(\mathbf{k}) \cdot \bigl( \partial_{k_\mu} \hat{\mathbf{d}}(\mathbf{k}) \times \partial_{k_\nu} \hat{\mathbf{d}}(\mathbf{k}) \bigr).\nonumber
\label{eq1}
\end{eqnarray}
The integration of Berry curvature $B_{xy}(\mathbf{k})$ over the Brillouin zone (BZ) gives the Chern number $C$ characterizing the topology of the lower band~\cite{Chern1946}.

We focus on the quantum metric and two global geometric quantities derived from it: the quantum volume and the gauge-invariant part of the Wannier spread functional. The quantum volume is defined as~\cite{Tomoki2021},
\begin{equation}
\mathrm{vol}_{g} = \int_{\mathrm{BZ}}  \sqrt{\det G(\mathbf{k})}\, d^2\mathbf{k},\label{eq3}
\end{equation}
where $\sqrt{\det G(\mathbf{k})}$ is the local quantum volume density.
The gauge-invariant part of the Wannier spread functional, also known as
the Marzari-Vanderbilt invariant~\cite{Nicola1997,Marzari2012,Tam2024}, is given by
\begin{equation}
\Omega_I = \frac{1}{(2\pi)^2}\int_{\mathrm{BZ}}\!\operatorname{Tr}G(\mathbf{k})\,d^2\mathbf{k},
\label{eq4}
\end{equation}
where the unit-cell area has been set to unity. For the eigenstate of a 2D two-band model, the quantum metric and Berry curvature satisfy the local geometric inequality~\cite{Roy2014,Peotta2015}: 
$\operatorname{Tr} G(\mathbf{k}) \geq 2\sqrt{\det G(\mathbf{k})}= |B_{xy}(\mathbf{k})|$.
Therefore, the vanishing of the Berry curvature coincides with the vanishing of the local volume element density. Integrating this inequality over the BZ yields a global inequality~\cite{Roy2014,Peotta2015,Tomoki2021,Mera2021},
\begin{equation}
2\pi^2\Omega_I \geq \mathrm{vol}_{g} \geq \pi |C| \label{global}.
\end{equation}

\begin{figure*}[htbp]
\centering
\includegraphics[width=0.8\textwidth]{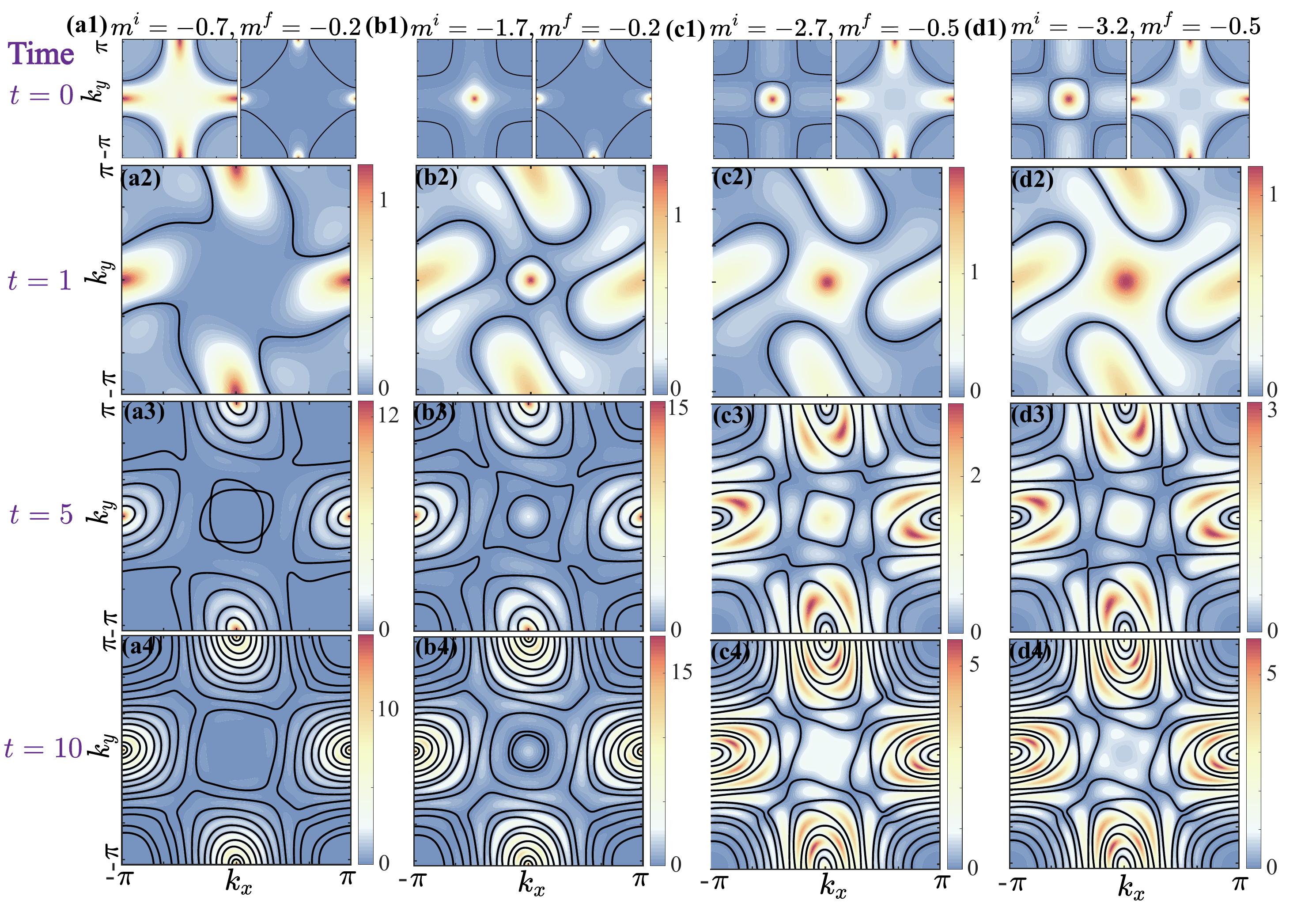}
\caption{Time evolution of the quantum-volume density $\sqrt{\det G(\mathbf{k},t)}$ in a quenched Chern insulator. The momentum-space distributions of $\sqrt{\det G(\mathbf{k},t)}$ in the BZ are shown at times $t=0,1,5,10$. Panels (a1)-(b4) correspond to quenches within the same topological phase, while panels (c1)-(d4) correspond to quenches across a topological phase boundary. In panels (a1)-(d1), the left and right panels show the ground-state quantum-volume density of the pre-quench and post-quench Hamiltonians, $H^{i}(\mathbf{k})$ and $H^{f}(\mathbf{k})$, respectively.  The quench parameters are 
(a1)-(a4) $m^i=-0.7$ and $m^f=-0.2$, (b1)-(b4) $m^i=-1.7$ and $m^f=-0.2$, (c1)-(c4) $m^i=-2.7$ and $m^f=-0.5$, (d1)-(d4) $m^i=-3.2$ and $m^f=-0.5$. In all panels, black contours indicate regions where $\det G(\mathbf{k},t)=0$.}
\label{Fig2}
\end{figure*}

As a concrete example, we consider the paradigmatic massive Dirac model~\cite{QI2006}, where
\begin{equation}
\mathbf{d}(\mathbf{k}) = (\sin k_x , \sin k_y , m + \cos k_x + \cos k_y),
\label{eqh}
\end{equation}
with the mass parameter $m$; the energy is measured in
units of the hopping strength. The energy bands are $\pm E({\bf k})=\pm \sqrt{\sin^2 k_x + \sin^2 k_y + (m + \cos k_x + \cos k_y)^2}$. As shown in Fig.~\ref{Fig1}, the system exhibits a trivial phase with $C=0$ for $|m|> 2$, a topological phase with $C=1$ for $-2<m <0$, and a topological phase with $C=-1$ for $0<m <2$. The phase transition occurs at $m =0, \pm 2$, where the energy gap closes. Using Eqs.~\eqref{eq3} and ~\eqref{eq4}, the quantum volume $\mathrm{vol}_g$ and Wannier spread $\Omega_I$ of the ground state are plotted as a function of the mass parameter $m$ [Fig.~\ref{Fig1}]. The global inequality (\ref{global}) is clearly seen. Note that both $\mathrm{vol}_g$ and $\Omega_I$ show discontinuities at the topological phase transition.
\begin{figure*}[htbp] 
\centering
\includegraphics[width=0.8\textwidth]{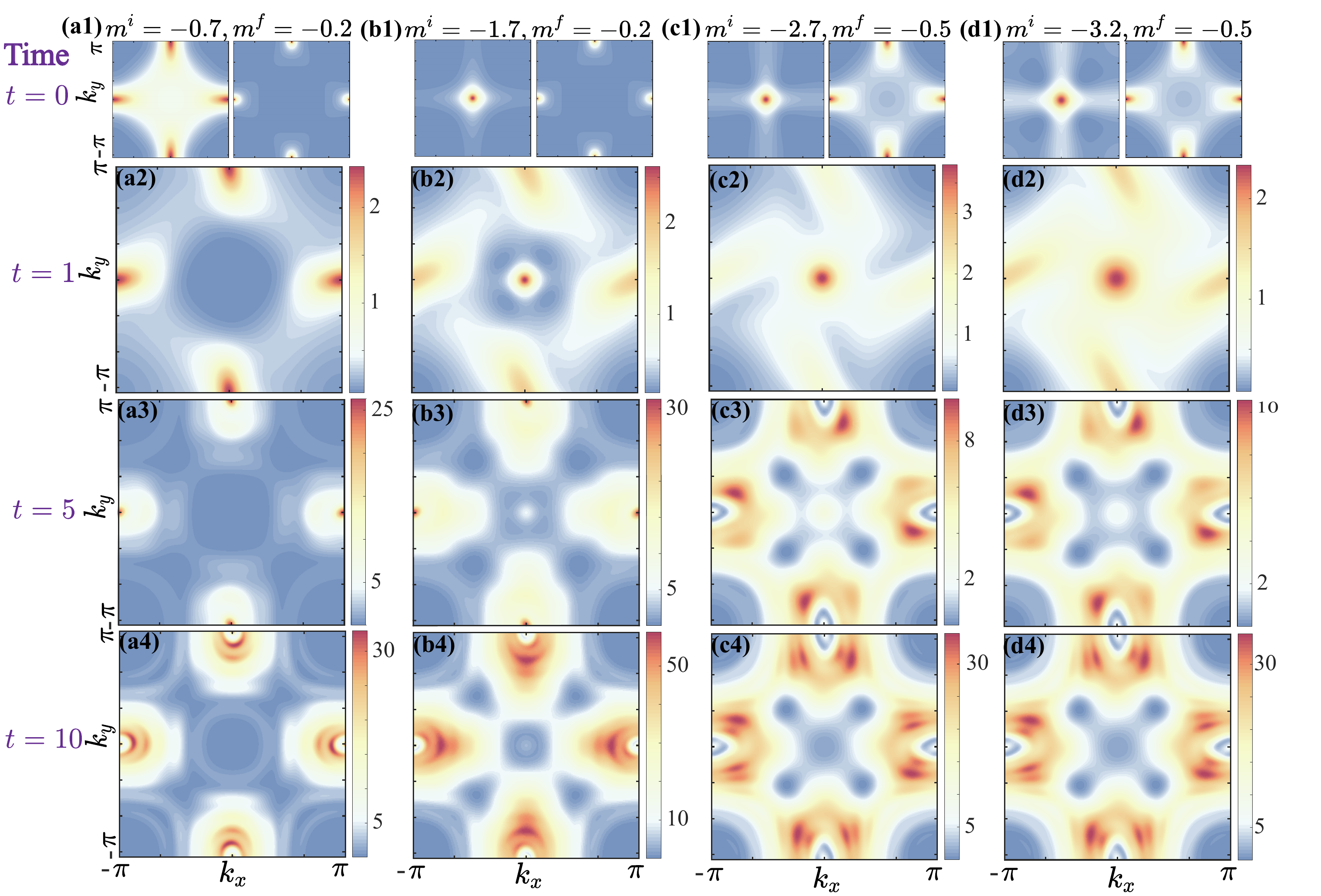}
\caption{The time evolution of the metric trace $\operatorname{Tr}G(\mathbf{k},t)$ corresponding to Fig.~\ref{Fig2}. Parameters are the same as those in corresponding panels in Fig.~\ref{Fig2}.}
\label{Fig3}
\end{figure*}

\section{Dynamical Quantum Geometry in a Quenched Chern insulator}\label{sec3}

We now consider a sudden quench of the massive Dirac model and investigate how quantum geometry is shaped by coherent nonequilibrium dynamics. The system is initially prepared in the ground state of $H^{i}(\mathbf{k}) = \mathbf{d}^{i}(\mathbf{k}) \cdot \boldsymbol{\sigma}$ in the form of Eq.~(\ref{eqh}) with the mass parameter $m^i$, before the Hamiltonian is instantaneously quenched to $H^{f}(\mathbf{k}) = \mathbf{d}^{f}(\mathbf{k}) \cdot \boldsymbol{\sigma}$ with $m^{f}$. We denote $\hat{\mathbf{d}}^{\alpha}(\mathbf{k}) = \mathbf{d}^{\alpha}(\mathbf{k})/{d}^{\alpha}(\mathbf{k})$ for $\alpha = i, f$. Assuming lattice translation invariance, the time-dependent density matrix at lattice momentum $\mathbf{k}$ can be written as
\begin{equation}\label{rhokt}
\rho_\mathbf{k}(t) = \frac{1}{2}\bigl[ \mathbb{I} - \hat{\mathbf{n}}(\mathbf{k},t)\cdot\boldsymbol{\sigma} \bigr],
\end{equation}
where the vector $\hat{\mathbf{n}}(\mathbf{k},t)$ describes the polarization of $\rho_\mathbf{k}(t)$ on the Bloch sphere. Under coherent evolution, straightforward calculations yield~\cite{SM1} 
\begin{eqnarray}
\hat{\mathbf{n}}(\mathbf{k},t) &=& \hat{\mathbf{d}}^{i}(\mathbf{k})\cos(2 d^{f} t) 
- \bigl[\hat{\mathbf{d}}^{i}(\mathbf{k}) \times \hat{\mathbf{d}}^{f}(\mathbf{k})\bigr] \sin(2 d^{f} t)\nonumber \\
\quad &+& 2 \hat{\mathbf{d}}^{f}(\mathbf{k})\,\bigl[\hat{\mathbf{d}}^{f}(\mathbf{k})\cdot \hat{\mathbf{d}}^{i}(\mathbf{k})\bigr] \sin^{2}\!\bigl(d^{f} t\bigr).
\label{eq:n_evolution} 
\end{eqnarray}
Thus the instantaneous Berry curvature and quantum metric for the time-evolving state (\ref{rhokt}) can be written as~\cite{Alessio2015,Hu2016}
\begin{align}
B_{xy}(\mathbf{k},t)
  &= \frac{1}{2}\,\hat{\mathbf{n}}(\mathbf{k},t)\!\cdot\![\partial_{k_x}\hat{\mathbf{n}}(\mathbf{k},t)\times\partial_{k_y}\hat{\mathbf{n}}(\mathbf{k},t)],\label{Bt}\\
G_{\mu\nu}(\mathbf{k},t)
  &= \frac{1}{4}\,\partial_{k_\mu}\hat{\mathbf{n}}(\mathbf{k},t)\!\cdot\!\partial_{k_\nu}\hat{\mathbf{n}}(\mathbf{k},t). \label{Gt}
\end{align}

As is well known, the Chern number associated with the evolving wavefunction,
\begin{equation}
C(t)=\frac1{2\pi}\int_{\rm BZ}B_{xy}(\mathbf k,t)d^2\mathbf{k}=C(0),\label{Ct}
\end{equation}
remains exactly conserved under unitary evolution, even when the post-quench Hamiltonians belong to a different topological phase~\cite{Alessio2015,Hu2016,Caio2016,Wilson2016}. This conservation gives rise to the topological incompatibility between the instantaneous wavefunction and Hamiltonian. We therefore ask whether
the quantum metric similarly retains initial-state memory or is dynamically
reconstructed during coherent evolution.

To address this question, we consider quenches both within the same
topological phase and across a topological phase transition, and examine 
local scalar quantities of the quantum metric, namely the quantum-volume
density $\sqrt{\det G(\mathbf{k},t)}$ and the metric trace
$\operatorname{Tr}G(\mathbf{k},t)$, along with the global quantum volume $\mathrm{vol}_g(t)$ and the Wannier spread $\Omega_I(t)$:
\begin{eqnarray}
\mathrm{vol}_{g}(t) &=& \int_{\mathrm{BZ}} 
\sqrt{\det G(\mathbf{k},t)}d^2\mathbf{k},\label{volgt}\\
\Omega_I(t) &=&\frac{1}{(2\pi)^2}\int_{\mathrm{BZ}}\!\operatorname{Tr}G(\mathbf{k},t)d^2\mathbf{k}.\label{eq8}
\end{eqnarray}
In this section, we analyze the dynamics of these quantities, and then in Sec.~\ref{sec4}
identify the underlying directional structure of the quantum metric and its physical origin.

\subsection{Evolution of the Local Quantum Geometry}\label{local}

We first examine local quantities of the quantum metric. In Figs.~\ref{Fig2}(a1)-(b4), we compare the quantum-volume density $\sqrt{\det G(\mathbf k,t)}$ in two quenches within the same topological phase ($C=1$) that share the same final Hamiltonian ($m^f=-0.2$) but start from distinct initial states. We see that $\sqrt{\det G(\mathbf k,t)}$ evolves at long times toward the characteristic momentum-space distribution of the post-quench ground state (see also Supplementary Material~\cite{SM1}).

This reconstruction is particularly evident for the second protocol [Figs.~\ref{Fig2}(b1)-(b4)]. Initially, the quantum-volume density of the state prepared at $m^i=-1.7$ is concentrated  around the $(k_x, k_y)=(0,0)$ point in the BZ, which differs qualitatively from the equilibrium distribution of the post-quench Hamiltonian, where the dominant peaks are located near $(\pm\pi,0)$ and $(0,\pm\pi)$ [Fig.~\ref{Fig2}(b1)]. During the subsequent coherent evolution [Figs.~\ref{Fig2}(b2)-(b4)], the initial peak at $(0, 0)$ is progressively suppressed, while new maxima emerge around the $(\pm\pi,0)$ and $(0,\pm\pi)$ points. Similar reconstruction behavior is observed for quenches across a topological phase boundary [Figs.~\ref{Fig2}(c1)-(d4)]. 

The corresponding evolution of the metric trace $\operatorname{Tr}G(\mathbf{k},t)$ is shown in Fig.~\ref{Fig3} and Supplemental Material ~\cite{SM1}, both for quenches within the same phase and across a phase transition. Similarly, the long-time momentum-space pattern of $\operatorname{Tr}G(\mathbf{k},t)$ is dynamically reconstructed and acquires characteristic features of the post-quench Hamiltonian.

Thus, these scalar metric observables acquire characteristic long-time structures associated with the post-quench Hamiltonian, even for different initial states and quench protocols.

\begin{figure}[tp]
\centering
\includegraphics[width=\columnwidth]{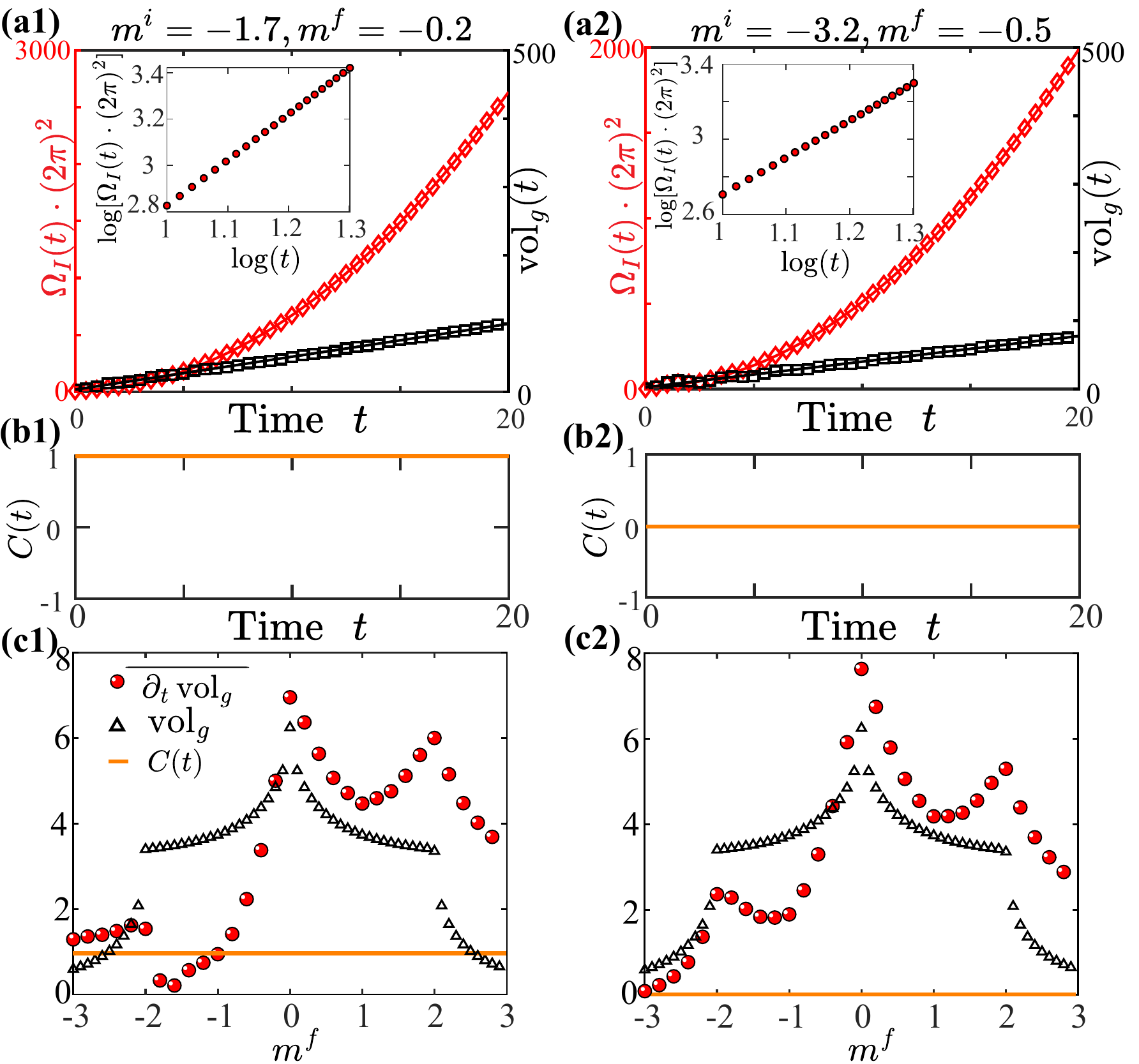}
\caption{Time evolution of global quantum geometry and the Chern number in a quenched Chern insulator. 
(a1)-(a2) Time evolution of the gauge-invariant Wannier spread functional $\Omega_I(t)$ (red diamonds) and quantum volume $\mathrm{vol}_g(t)$ (black squares), calculated from Eqs.~\eqref{volgt} and \eqref{eq8}, respectively. Solid lines denote fits using $\Omega_I/(2\pi)^2=at^2+bt+c$ (red) and $\mathrm{vol}_g=\alpha t+e$ (black). Insets show log-log plots of $\Omega_I/(2\pi)^2$ versus $t$. 
(b1)-(b2) Time-dependent Chern number $C(t)$ corresponding to (a1) and (a2), respectively. 
(c1)-(c2) $\overline{\alpha}=\overline{\partial_t\mathrm{vol}}_g$ as a function of the post-quench mass $m^f$. In (c1) and (c2), the initial state is same as (a1) and (a2), respectively. The $\overline{\mathrm{vol}}_g$ is the average over the time interval $t\in[10,20]$. The orange curve denotes the Chern number at time $t=20$ in the corresponding quench. Black curves show the quantum volume of the ground state of $H^f(\mathbf{k})$. }
\label{Fig4}
\end{figure}

\subsection{Evolution of the Global Quantum Geometry}

Having established the evolution of the local quantum geometry, we now investigate global quantities, i.e., quantum volume $\mathrm{vol}_g(t)$ and the Wannier spread $\Omega_I(t)$~\cite{SM1}. Figures~\ref{Fig4}(a1) and (a2) 
illustrate their evolution for representative quenches within the same topological phase [Fig.~\ref{Fig4}(a1)] and across a topological phase boundary [Fig.~\ref{Fig4}(a2)]. Despite the different quench protocols, three generic features are observed:

(1) $\Omega_I(t)$
and
$\mathrm{vol}_g(t)$ exhibit distinct dynamical scalings at long times, 
\begin{equation}\label{scaling}
\Omega_I(t)\propto t^2, \hspace{1mm}\mathrm{vol}_g(t)\propto t.
\end{equation}
These long-time scalings are irrespective of the initial state specifics [Figs.~\ref{Fig4}(a1) and (a2)]. This is in sharp contrast to the time-dependent Chern number $C(t)$ in Eq.~(\ref{Ct}), which is conserved in time under unitary evolution [Figs.~\ref{Fig4}(b1) and (b2)]. Thus topological invariant, the quantum volume, and Wannier spread exhibit three distinct dynamical scalings in coherent evolution. Note that the scaling (\ref{scaling}) has been recently shown in 1D lattice systems~\cite{Wu2024,Tang2026}.

Note the $t^2$ scaling of the Wannier spread can be understood from the well-known expansion~\cite{Blount1962,Resta1994,Vanderbilt2018}: $
G_{\nu\nu}(t)= \text{Var}(\hat{x}) + 2t \, \text{Cov}(\hat{x}, \hat{v}_\nu) + t^2 \, \text{Var}(\hat{v}_\nu)$ for $\nu \in \{x,y\}$,  where $\partial_t \hat{x} = i[H^{f}, \hat{x}] = \hat{v}_\nu$ defines the velocity operator $\hat{v}_\nu$, $\text{Var}(\hat{A}) \equiv \langle \hat{A}^2 \rangle - \langle \hat{A} \rangle^2$ and $\text{Cov}(\hat{A},\hat{B}) \equiv \tfrac{1}{2}\langle \{\hat{A},\hat{B}\} \rangle - \langle \hat{A} \rangle\langle \hat{B} \rangle$ are evaluated over the initial state. Thus at long times, the quadratic contribution from the velocity variance dominates. 

(2) The global geometric inequality remains satisfied throughout the coherent evolution [Figures~\ref{Fig4}(a1) and (a2)],
\begin{equation}
2\pi^2\Omega_I(t)>\mathrm{vol}_g(t)>\pi |C(t)|.
\end{equation}
Different from the equilibrium inequality (\ref{global}), we expect its lower bound generally no longer saturated due to distinct dynamical scalings of $\Omega_I(t)$, $\mathrm{vol}_g(t)$ and $C(t)$.

(3) The long-time global quantum geometry retains characteristic features of the post-quench Hamiltonian. In Figs.~\ref{Fig4}(c1) and (c2), we fix the initial state and vary the post-quench Hamiltonian by tuning the final mass parameter $m^f$; for each $m^f$, we extract from $\mathrm{vol}_g=\alpha t+e$ the growth rate $\alpha$ and plot it as a function of $m^f$ (red curves). As shown, $\alpha$ exhibits nonanalytic features near $m^f=0,\pm2$, resembling the behavior of the ground-state quantum volume of the corresponding post-quench Hamiltonian (black curves). This sharply contrasts with the time-averaged Chern number (yellow curve), whose value is always pinned by the initial state.

\section{Directional dynamical reconstruction of quantum geometry}
\label{sec4}

The distinct long-time scaling behaviors of the quantum volume and the
Wannier spread in Eq.~(\ref{scaling}) indicate that the quantum metric does
not evolve isotropically after the quench. To uncover the underlying local
structure, we perform the spectral decomposition of the instantaneous
quantum metric,
\begin{equation}\label{Gkt12}
G(\mathbf{k},t)
=
\lambda_1(\mathbf{k},t)
v_1(\mathbf{k},t)v_1^{T}(\mathbf{k},t)
+
\lambda_2(\mathbf{k},t)
v_2(\mathbf{k},t)v_2^{T}(\mathbf{k},t),
\end{equation}
where $\lambda_1\geq\lambda_2$ are the ordered eigenvalues and
$v_{1,2}$ are the corresponding principal axes of the quantum distance.

\begin{figure*}[tp]
\centering
\includegraphics[width=0.65\textwidth]{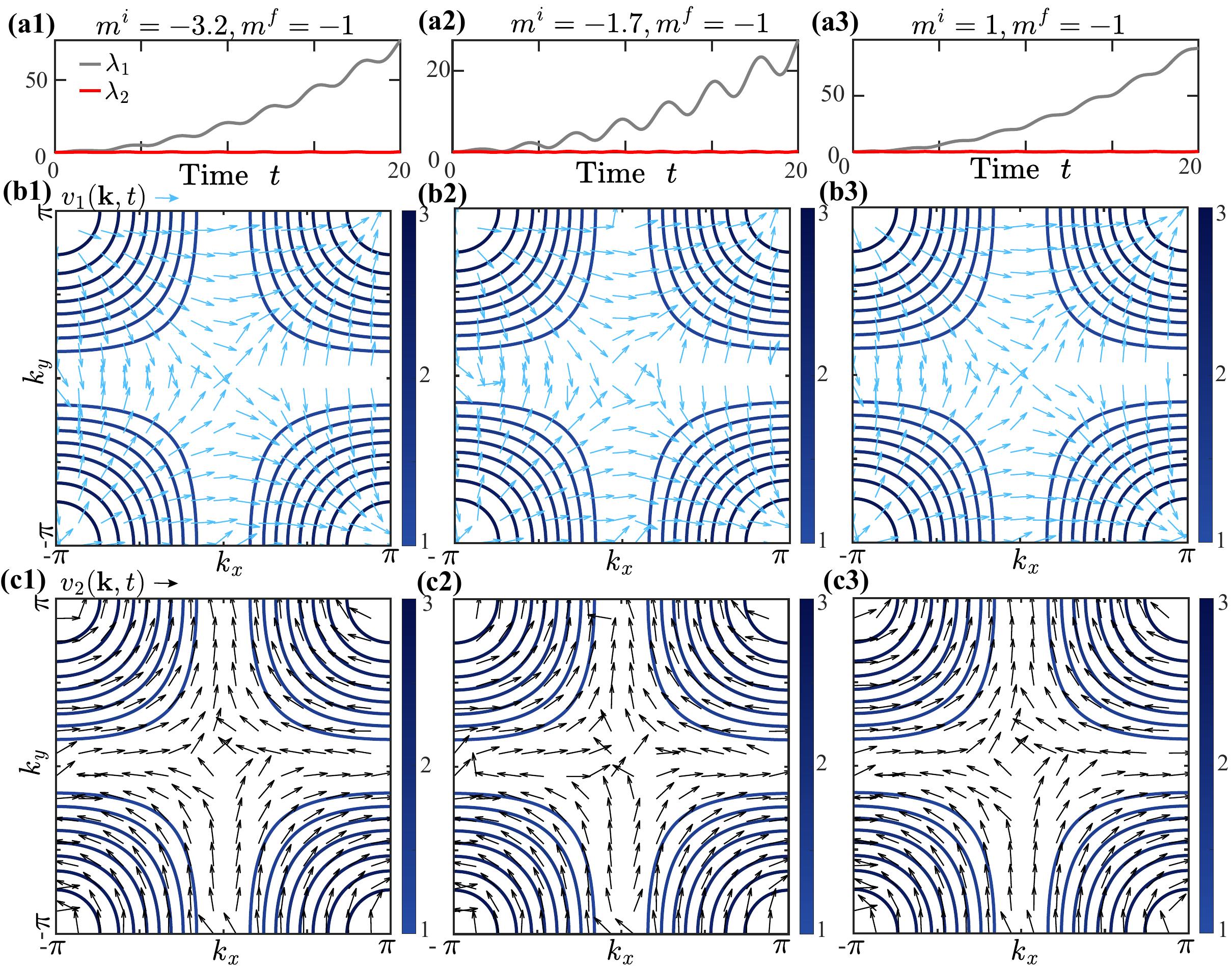}
\caption{Anisotropic dynamical reconstruction of quantum geometry. The three columns respectively correspond to three distinct quenches, which share identical $m^f=-1$ but (a1) $m^i=-3.2$, (a2) $m^i=-1.7$, and (a3) $m^i=1$.
(a1)-(a3) Time evolution of the metric eigenvalues $\lambda_{1,2}(\mathbf{k},t)$ at $\mathbf{k}=(\pi/5,\pi/2)$. (b1)-(b3) Long-time ($t=20$) distributions of the principal axis $v_1$ associated with the larger metric eigenvalue $\lambda_1$ [c.f. Eq.~(\ref{Gkt12})]. Arrows indicate the orientation $(\cos(\theta_{\mathbf{k}}(t)/2),\sin(\theta_{\mathbf{k}}(t)/2))$. (c1)-(c3) Corresponding distributions of the principal axis $v_2$ at $t=20$ associated with the smaller metric eigenvalue $\lambda_2$. Arrows indicate the orientation $(-\sin(\theta_{\mathbf{k}}(t)/2),\cos(\theta_{\mathbf{k}}(t)/2))$. respectively. In (b1)-(c3), blue curves denote the constant-energy contours of the post-quench Hamiltonian $H^f(\mathbf{k})$.}
\label{Fig5}
\end{figure*}

As shown in Figs.~\ref{Fig5}(a1) and (a2), the two eigenvalues exhibit
qualitatively different dynamical behaviors. The larger eigenvalue obeys
\begin{equation}
\lambda_1(\mathbf{k},t)\propto t^2 ,
\label{lambda1}
\end{equation}
whereas the smaller eigenvalue remains nearly unchanged during the
evolution,
\begin{equation}
\lambda_2(\mathbf{k},t)\simeq
\lambda_2(\mathbf{k},0).
\label{lambda2}
\end{equation}
This increasingly anisotropic local metric explains the distinct global
scalings: $\Omega_I(t)\sim \operatorname{Tr}G(\mathbf k,t)\sim t^2$ and
$\mathrm{vol}_{g}(t)\sim\sqrt{\det G(\mathbf k,t)}\sim t$. Coherent
evolution thus selectively enhances quantum distinguishability along one
principal direction while leaving the orthogonal direction largely unaffected.

To identify the origin of this directional reorganization, we separate the
contribution generated by coherent evolution from that inherited from the
initial state. In the eigenbasis of the post-quench Hamiltonian $H^f(\mathbf{k})$, the
time-evolved state can be written as
\begin{equation}
|\psi(\mathbf{k},t)\rangle
=
g_{\mathbf{k}}
|u_-^f(\mathbf{k})\rangle
+
e^{-i\phi_{\mathbf{k}}(t)}
e_{\mathbf{k}}
|u_+^f(\mathbf{k})\rangle,
\label{psit}
\end{equation}
where $|u_\pm^f(\mathbf{k})\rangle$ are eigenstates of the post-quench
Hamiltonian and $g_{\mathbf{k}}$ and $e_{\mathbf{k}}$ are amplitudes set by
the initial state. The relative dynamical phase
$\phi_{\mathbf{k}}(t)=2E_f(\mathbf{k})t$ contains the only explicit time
dependence. To isolate its contribution, we consider the quantum distance
$ds_{\phi}^{2}$ between neighboring states arising solely from the phase
difference $d\phi_{\mathbf{k}}$. Equation~(\ref{psit}) gives
\begin{eqnarray}\label{normal}
ds_{\phi}^{2}
&\simeq&
|g_{\mathbf{k}}|^2|e_{\mathbf{k}}|^2
(d\phi_{\mathbf{k}})^2\nonumber\\
&=&4|g_{\mathbf{k}}|^2
|e_{\mathbf{k}}|^2
|\nabla_{\mathbf{k}}E_f(\mathbf{k})|^2
t^2
dk_{\perp}^{2}.
\end{eqnarray}
In the second line, the momentum displacement is decomposed as
$d\mathbf{k}=dk_{\perp}\hat{\mathbf e}_{\perp}+dk_{\parallel}\hat{\mathbf e}_{\parallel}$,
where $\hat{\mathbf e}_{\perp}$ and $\hat{\mathbf e}_{\parallel}$ are normal
and tangent to the constant-energy contours of $H^f(\mathbf{k})$, respectively.

Equation~(\ref{normal}) makes the directional origin of the growing metric
component explicit. Along $\hat{\mathbf e}_{\perp}$, neighboring momentum
states have different post-quench energies and hence accumulate different
dynamical phases. Their growing phase separation enhances quantum
distinguishability and produces $\lambda_1$. Along
$\hat{\mathbf e}_{\parallel}$, the energy difference vanishes to leading
order, so neighboring states remain phase locked. Therefore, essentially no growing quantum distance
is generated in this direction, and the corresponding component $\lambda_2$
remains set primarily by the initial geometry.

To determine the orientation of the principal axes of the time-evolving quantum metric, we parameterize the real symmetric matrix $G(\mathbf{k},t)$ as
\begin{equation}
G(\mathbf{k},t)=\beta_{\mathbf{k}}(t)\mathbb{I}
+\frac{\Delta_{\mathbf{k}}(t)}{2}
\left[\sin\theta_{\mathbf{k}}(t)\sigma_x
+\cos\theta_{\mathbf{k}}(t)\sigma_z\right],
\end{equation}
where $\beta_{\mathbf{k}}(t) = \frac{G_{xx}(\mathbf{k},t) + G_{yy}(\mathbf{k},t)}{2}$ and $\Delta_{\mathbf{k}}(t) = \sqrt{\left[ G_{xx}(\mathbf{k},t) - G_{yy}(\mathbf{k},t) \right]^2 + 4 G_{xy}^2(\mathbf{k},t)}$, with $\sin\theta_{\mathbf{k}}(t) = \frac{2 G_{xy}(\mathbf{k},t)}{\Delta_{\mathbf{k}}(t)}$ and $\cos\theta_{\mathbf{k}}(t) = \frac{G_{xx}(\mathbf{k},t) - G_{yy}(\mathbf{k},t)}{\Delta_{\mathbf{k}}(t)}$. The corresponding normalized eigenvectors are

\begin{equation}\label{eq:eigenG}
v_{1}(\mathbf{k},t)=
\begin{pmatrix}
\cos\left( \theta_{\mathbf{k}}(t)/2 \right) \\
\sin\left( \theta_{\mathbf{k}}(t)/2 \right)
\end{pmatrix},
\quad
v_{2}(\mathbf{k},t)=
\begin{pmatrix}
-\sin\left( \theta_{\mathbf{k}}(t)/2 \right) \\
\cos\left( \theta_{\mathbf{k}}(t)/2 \right)
\end{pmatrix}.
\end{equation}
Figures~\ref{Fig5}(b1)--(c3) show the momentum-space distributions of these
principal axes at long times for distinct initial states evolving under the
same post-quench Hamiltonian. In all cases, the amplified axis $v_1$ aligns with the normal
direction of the post-quench constant-energy contours, whereas the nearly
inert axis $v_2$ aligns with their tangent direction, in agreement with the
dynamical-phase argument above.

\begin{figure}[tb]
\centering
\includegraphics[width=1\columnwidth]{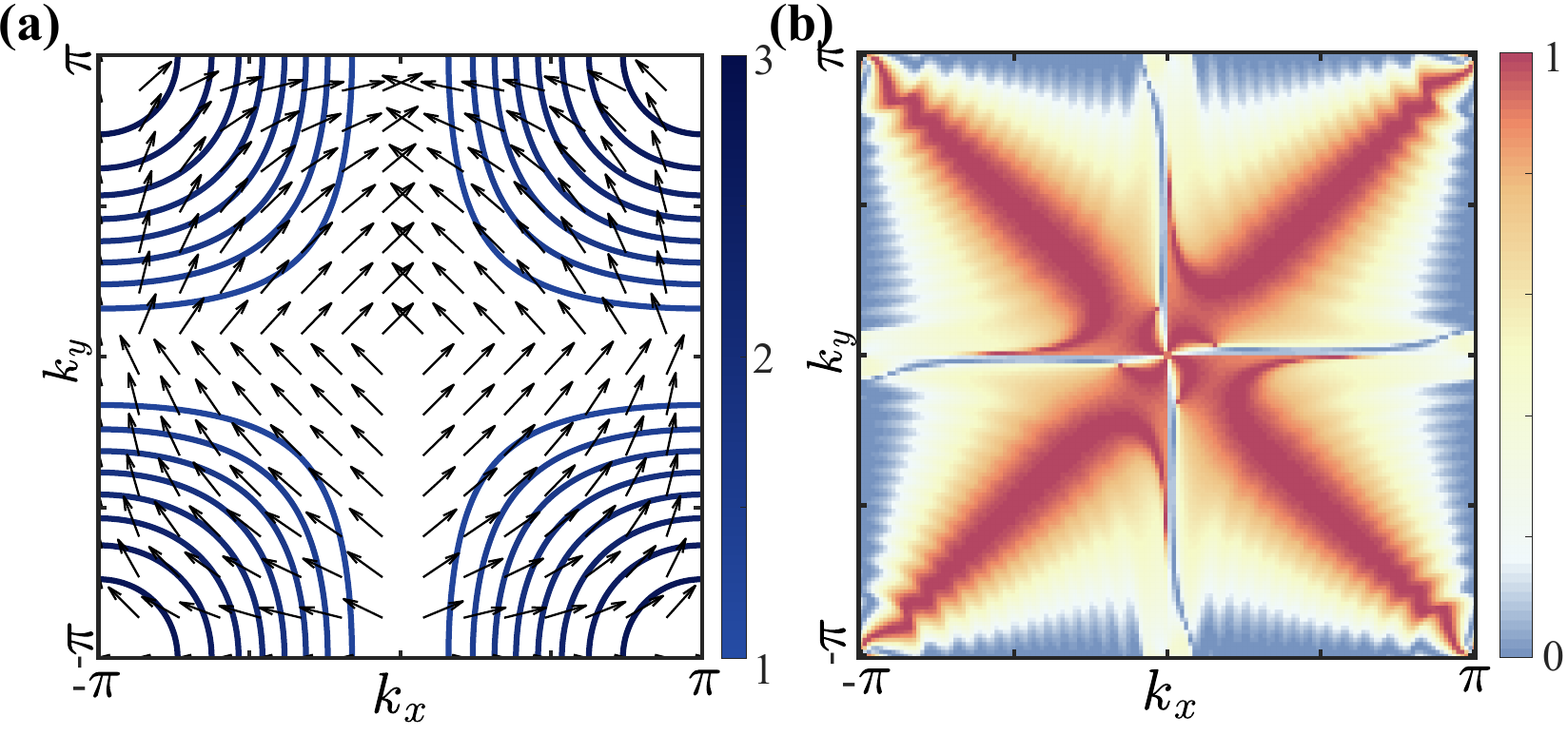}
\caption{Comparison between the principal axes of the static and dynamical quantum metrics. (a) Smaller principal axis $v_2^f$ of the ground-state quantum metric of $H^f(\mathbf{k})$ with $m^f=-1$. Blue curves denote constant-energy contours of $H^f(\mathbf{k})$. (b) Momentum-resolved overlap $|\langle v_2(\mathbf{k},t)|v_2^f(\mathbf{k})\rangle|^2$ between the smaller principal axis $v_2$ of the time-evolving metric at $t=20$ and that of the ground-state metric of $H^f(\mathbf{k})$. Here, we take the same $v_2(\mathbf{k},t)$ as Fig.~\ref{Fig5}(c1), corresponding to the metric at $t=20$ in a quench from $m^i=-3.2$ to $m^f=-1$.}
\label{Fig6}
\end{figure}

This anisotropic reorientation is a genuinely dynamical effect with no counterpart in the static ground-state geometry of the post-quench Hamiltonian $H^f(\mathbf{k})$. Figure~\ref{Fig6}(a) shows that the smaller principal axis $v_2^f$ of the ground-state metric does not generally follow the tangent direction of the constant-energy contours. Further results are presented in the Supplementary Material~\cite{SM1}. Moreover, the momentum-resolved overlap between the long-time axis $v_2(\mathbf{k},t)$ of the evolving state and $v_2^f$ varies strongly across the BZ [Fig.~\ref{Fig6}(b)], with substantial regions of small overlap. Thus, the alignment of the dynamical metric axes with the energy-gradient and equal-energy directions is not inherited from the static geometry of $H^f(\mathbf{k})$, but is selected during coherent evolution.

\section{Conclusion}\label{sec5}

In this work, we have investigated the quantum geometry of coherently evolving states following a sudden quench in a two-dimensional Chern insulator. While the Chern number remains conserved under unitary dynamics, scalar observables of the quantum metric develop characteristic long-time momentum-space structures associated with the post-quench Hamiltonian. Globally, the quantum volume and gauge-invariant Wannier spread exhibit distinct long-time scalings, in contrast to the conserved Chern number.

Most importantly, we have identified the local structure underlying these behaviors. In the 2D quenched Chern insulators considered here, coherent evolution dynamically selects a principal frame of the quantum metric: the growing component aligns with the energy-gradient direction of the post-quench Hamiltonian, while the nearly inert component follows the tangent direction of its constant-energy contours and retains memory of the initial geometry. This anisotropic reorientation originates from momentum-dependent relative dynamical phases. It is distinct from the principal-axis structure of the static post-quench ground-state metric and therefore has no static counterpart. Our results demonstrate a dynamical anisotropic reconstruction of the local quantum metric and suggest possibilities for controlling quantum geometry out of equilibrium. Although demonstrated here for the massive Dirac model, the underlying mechanism is expected to extend to other high-dimensional systems with coherent multiband dynamics.

The predicted dynamical quantum geometry is accessible with current quantum simulation platforms. For instance, in ultracold atoms, the quench protocol can be readily realized in optical-lattice implementations of massive Dirac or Haldane models~\cite{Jotzu2014}, while the momentum-resolved quantum metric can be reconstructed through quantum-state tomography~\cite{Yi2023} of the quantum geometric tensor.

\section*{Acknowledgments}

Z.X.L acknowledges support by the National Natural Science Foundation of China (Grant No. 12574301) and the Zhejiang Provincial Natural Science Foundation of China under Grant No. LZ25A040004. This work was supported by the Key Project of the National Natural Science Foundation of China Joint Funds (Grants No.~U25A20197), and the National Natural Science Foundation of China (Grants No. 12374246).


\begin{thebibliography}{99}



\bibitem{Torma2023} P. Törmä, Essay: Where can quantum geometry lead us?, Phys. Rev. Lett. \textbf{131}, 240001 (2023).
\bibitem{Liu2025} T. Liu, X. Qiang, H. Lu, and X. C. Xie, Quantum geometry in condensed matter, Natl. Sci. Rev. \textbf{12}, nwae334 (2025).
\bibitem{Yu2025} J. Yu, B. A. Bernevig, R. Queiroz, E. Rossi, P. Törmä, and B. Yang, Quantum geometry in quantum materials, npj Quantum Mater. \textbf{10}, 101 (2025).

\bibitem{Provost1980} J. P. Provost and G. Vallee, Riemannian structure on manifolds of quantum states, Commun. Math. Phys. \textbf{76}, 289 (1980).
\bibitem{Ma2010} Y. Ma, S. Chen, H. Fan, and W. Liu, Abelian and non-Abelian quantum geometric tensor, Phys. Rev. B \textbf{81}, 245129 (2010).

\bibitem{Berry1984} M. V. Berry, Quantal phase factors accompanying adiabatic changes, Proc. R. Soc. A \textbf{392}, 45 (1984).
\bibitem{Xiao2010} D. Xiao, M. Chang, and Q. Niu, Berry phase effects on electronic properties, Rev. Mod. Phys. \textbf{82}, 1959 (2010).


\bibitem{Fubini1904} G. Fubini, Sulle metriche definite da una forma hermitiana, Atti del Reale istituto Veneto di Scienze, Lettere ed Arti \textbf{63}, 502 (1904).
\bibitem{Study1905} E. Study,  Kürzeste Wege im komplexen Gebiet, Math. Ann. \textbf{60}, 321 (1905).


\bibitem{Peotta2015} S. Peotta and P. Törmä, Superfluidity in topologically nontrivial flat bands, Nat. Commun. \textbf{6}, 8944 (2015).
\bibitem{Hu2019}  X. Hu, T. Hyart, D. I. Pikulin, and E. Rossi, Geometric and conventional contribution to the superfluid weight in twisted bilayer graphene, Phys. Rev. Lett. \textbf{123}, 237002 (2019).
\bibitem{Xie2020}  F. Xie, Z. Song, B. Lian, and B. A. Bernevig,  Topology-bounded superfluid weight in twisted bilayer graphene, Phys. Rev. Lett. \textbf{124}, 167002 (2020).
\bibitem{Julku2021} A. Julku, G. M. Bruun, and P. Törmä, Quantum geometry and flat band Bose-Einstein condensation, Phys. Rev. Lett. \textbf{127}, 170404 (2021).
\bibitem{Ahn2022} J. Ahn, G. Guo, N. Nagaosa, and A. Vishwanath, Riemannian geometry of resonant optical responses, Nat. Phys. \textbf{18}, 290 (2022).
\bibitem{Li2026} Y. Li and C. Liu, Quantum-metric-based optical selection rules, Phys. Rev. Lett. \textbf{136}, 046901 (2026).
\bibitem{Roy2014} R. Roy, Band geometry of fractional topological insulators, Phys. Rev. B \textbf{90}, 165139 (2014).
\bibitem{WuL2024} A. Wu, S. Sarkar, X. Wan, K. Sun, and S. Lin, Quantum-metric-induced quantum Hall conductance inversion and reentrant transition in fractional Chern insulators, Phys. Rev. Res. \textbf{6}, L032063 (2024).
\bibitem{Shavit2024} G. Shavit and Y. Oreg, Quantum geometry and stabilization of fractional Chern insulators far from the ideal limit, Phys. Rev. Lett. \textbf{133}, 156504 (2024).
\bibitem{Hu2022} X. Hu, T. Hyart, D. I. Pikulin, and E. Rossi, Quantum-metric-enabled exciton condensate in double twisted bilayer graphene, Phys. Rev. B \textbf{105}, L140506 (2022).

 
\bibitem{Asteria2019} L. Asteria, D. T. Tran, T. Ozawa, M. Tarnowski, B. S. Rem, N. Fläschner, K. Sengstock, N. Goldman, and C. Weitenberg, Measuring quantized circular dichroism in ultracold topological matter, Nat. Phys. \textbf{15}, 449 (2019).
\bibitem{Tan2019} X. Tan, D. Zhang, Z. Yang, J. Chu, Y. Zhu, D. Li, X. Yang, S. Song, Z. Han, Z. Li \textit{et al.}, Experimental measurement of the quantum metric tensor and related topological phase transition with a superconducting qubit, Phys. Rev. Lett. \textbf{122}, 210401 (2019).
\bibitem{Yu2020} M. Yu, P. Yang, M. Gong, Q. Cao, Q. Lu, H. Liu, S. Zhang, M. B. Plenio, F. Jelezko, T. Ozawa \textit{et al.}, Experimental measurement of the quantum geometric tensor using coupled qubits in diamond, Natl. Sci. Rev. \textbf{7}, 254 (2020).
\bibitem{Kim2025} S. Kim, Y. Chung, Y. Qian, S. Park, C. Jozwiak, E. Rotenberg, A. Bostwick, K. S. Kim, and B. Yang, Direct measurement of the quantum metric tensor in solids, Science \textbf{388}, 1050 (2025).
\bibitem{Zhao2025} T. Zhao, A. Wang, Z. Zhang, Z. Cao, X. Liu, and Z. Liao, Magnetic field induced quantum metric dipole in Dirac semimetal $Cd_3As_2$, Phys. Rev. Lett. \textbf{135}, 026601 (2025).

\bibitem{Davide2020}  D. Rattacaso, P. Vitale, and A. Hamma, Quantum geometric tensor away from equilibrium,  J. Phys. Commun. \textbf{4}, 055017 (2020).
\bibitem{Wu2024} S. Wu, Z. Guo, Z. Xiong, Y. Yao, and Z. Cai, Universal critical dynamics of quantum geometry, arXiv:2401.17885 [cond-mat.quant-gas].
\bibitem{Tang2026} J. Tang, X. Hou, Y. Huang, H. Guo, and C. Chien, Evolution of quantum geometric tensor of one-dimensional periodic systems after a quench, Phys. Rev. B \textbf{113}, 174301 (2026).

\bibitem{Alessio2015} L. D'Alessio and M. Rigol, Dynamical preparation of Floquet Chern insulators, Nat. Commun. \textbf{6}, 8336 (2015).
\bibitem{Hu2016} Y. Hu, P. Zoller, and J. C. Budich, Dynamical buildup of a quantized Hall response from nontopological states, Phys. Rev. Lett. \textbf{117}, 126803 (2016).
\bibitem{Caio2016} M. D. Caio, N. R. Cooper, and M. J. Bhaseen, Hall response and edge current dynamics in Chern insulators out of equilibrium, Phys. Rev. B \textbf{94}, 155104 (2016).
\bibitem{Wilson2016} J. H. Wilson, J. C. W. Song, and G. Refael, Remnant geometric Hall response in a quantum quench, Phys. Rev. Lett. \textbf{117}, 235302 (2016).


\bibitem{Sun2018} W. Sun, C. Yi, B. Wang, W. Zhang, B. C. Sanders, X. Xu, Z. Wang, J. Schmiedmayer, Y. Deng, X. Liu \textit{et al.}, Uncover topology by quantum quench dynamics, Phys. Rev. Lett. \textbf{121}, 250403 (2018).



\bibitem{Chern1946} S. Chern, Characteristic classes of Hermitian manifolds, Ann. Math. \textbf{47}, 85 (1946).



\bibitem{Tomoki2021} T. Ozawa and B. Mera, Relations between topology and the quantum metric for Chern insulators, Phys. Rev. B \textbf{104}, 045103 (2021).
\bibitem{Nicola1997} N. Marzari and D. Vanderbilt, Maximally localized generalized Wannier functions for composite energy bands, Phys. Rev. B \textbf{56}, 12847 (1997).
\bibitem{Marzari2012} N. Marzari, A. A. Mostofi, J. R. Yates, I. Souza, and D. Vanderbilt, Maximally localized Wannier functions: Theory and applications, Rev. Mod. Phys. \textbf{84}, 1419 (2012).
\bibitem{Tam2024} P. M. Tam, J. Herzog-Arbeitman, and J. Yu, Corner charge fluctuation as an observable for quantum geometry and entanglement in two-dimensional insulators, Phys. Rev. Lett. \textbf{133}, 246603 (2024).


\bibitem{Mera2021} B. Mera and T. Ozawa, Kähler geometry and Chern insulators: Relations between topology and the quantum metric, Phys. Rev. B \textbf{104}, 045104 (2021).


\bibitem{QI2006} X. Qi, Y. Wu, and S. Zhang, Topological quantization of the spin Hall effect in two-dimensional paramagnetic semiconductors, Phys. Rev. B \textbf{74}, 085308 (2006).


\bibitem{SM1} See Supplemental Material for a detailed derivation of the instantaneous quantum metric, and  additional figures (Figs.~S1-S4) showing the same quantum geometric  quantities and their momentum-space distributions as in Figs.~2-6, for different quench parameters. 



\bibitem{Blount1962}  E. I. Blount, Formalisms of band theory, Solid State Phys. \textbf{13}, 305 (1962).
\bibitem{Resta1994} R. Resta, Macroscopic polarization in crystalline dielectrics: The geometric phase approach, Rev. Mod. Phys. \textbf{66}, 899 (1994).

\bibitem{Vanderbilt2018} D. Vanderbilt, \textit{Berry Phases in Electronic Structure Theory: Electric Polarization, Orbital Magnetization and Topological Insulators} (Cambridge University Press, Cambridge, UK, 2018).

\bibitem{Jotzu2014} G. Jotzu, M. Messer, R. Desbuquois, M. Lebrat, T. Uehlinger, D. Greif, and T. Esslinger, Experimental realization of the topological Haldane model with ultracold fermions, Nature (London) \textbf{515}, 237 (2014).
\bibitem{Yi2023} C. Yi, J. Yu, H. Yuan, R. Jiao, Y. Yang, X. Jiang, J. Zhang, S. Chen, and J. Pan, Extracting the quantum geometric tensor of an optical Raman lattice by Bloch-state tomography, Phys. Rev. Res. \textbf{5}, L032016 (2023).






\end{thebibliography}
\end{document}